# Machine Learning in Materials Modeling – Fundamentals and the Opportunities in 2D Materials

**Shreeja Das[1], Hansraj Pegu[1], Kisor Sahu[1], Ameeya Kumar Nayak[2], Seeram Ramakrishna[3], Dibakar Datta[4], Soumya Swayamjyoti[1,5,*]**

[1] School of Minerals, Metallurgical and Materials Engineering, Indian Institute of Technology (IIT) Bhubaneswar, India.

[2] Department of Mathematics, Indian Institute of Technology (IIT) Roorkee, India.

[3] Department of Mechanical Engineering, National University of Singapore (NUS), Singapore.

[4] Department of Mechanical and Industrial Engineering, Newark College of Engineering, New Jersey Institute of Technology, USA.

[5] NetTantra Technologies (India) Pvt. Ltd., Bhubaneswar, India

[*]**Corresponding Author**
Soumya Swayamjyoti (swayam.sj@gmail.com)

## 1 The Launch Platform for Machine Learning

The entire progression of human society is associated with the development or discovery of newer materials with newer and exciting functionalities. Fig. 1 attempts to capture this in a very simplistic schematic.

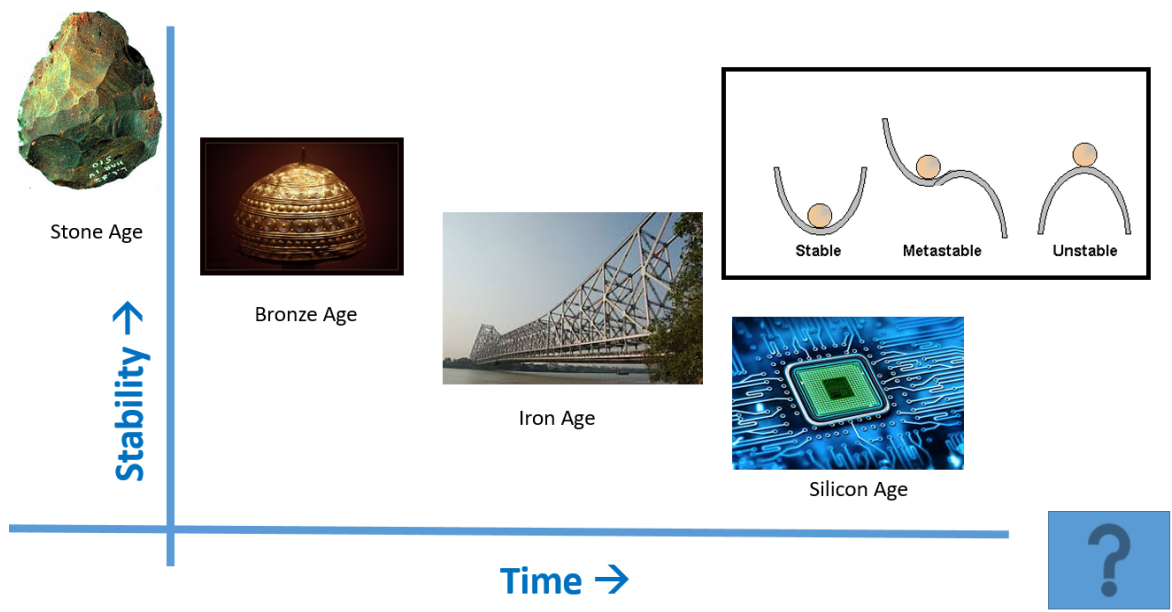

**Fig. 1:** Different stages of human civilizations: The stone, Bronze, Iron/steel and Silicon ages. During the progression, we learned to use materials which are further from their stable forms. Concept of the stability is schematically represented in the inset. The next materials to shape our future is not clear at the moment. Many believe that it is probably based on 2-dimensional materials (such as graphene, hexagonal boron nitride, transition metal dichalcogenides, etc. and heterostructures formed by them).

At the very dawn of human era, our ancestors learned to use stones to survive against bigger and stronger predators. It was also very useful against animals which were smaller but extremely dangerous, because of their possession of chemical weapons in their inventories, for example snakes and other lizards that could use their venom with fatal consequences. We will not really classify that time as the onset of human civilization because they really did not have to discover stones – because it was almost everywhere; they had to just find robust and sharper ones for better performance. It is nearly universally accepted that the true onset of human civilization started with the learning of the usage of fire; it gave them distinctive capabilities that truly differentiated them from any other animal on the planet. Strangely enough, this capability gap could not be breached by any other animal species even today. With the learning of usage of fire, humans could discover the first engineered materials – metals. Probably, copper was the first metal that was discovered by human, but it was not as useful as its alloyed counterpart: Bronze. Although at that time they did not understand much about alloying as we do now; nevertheless, it will be naïve to believe that they did not realize the unique properties of this newly discovered metal. The next big leap in human civilization had to wait for a millennia, until they could really master the technology of raising the temperature in a controlled environment. This will allow them the discovery of iron. Soon they will realize, like the previous instance, that it is not as useful as its alloy – steel. The far superior properties of this new material coupled with its relative abundance leading to its cheap prices will usher the industrial revolution. This is an era of unprecedented growth, unseen and unparalleled in the history of mankind, that will change the fate of both human civilization as well as that of the planet. Because of this accelerated pace of development of human civilization, the next big leap did not have to wait for longer; it came in a time span shorter than even half a millennium. This big leap led us to the information era, which has a very intimate connection to the heart and soul of this book. The discovery of the material that led us to this era is the use of silicon-based semiconductors. But this not just about a new material alone, but also of a new technological paradigm. Let us explore this in a bit detail. Prior to this era, we used to achieve complex functionality in a machine through 'assembly' of many different simple parts. However, in this new technology regime, functionality is achieved by 'embedding' the desired property in material itself. We will contrast the big difference between 'assembly' vs. 'embedding' through an example. A physical electrical switch (an ordinary switch that we use in our houses to switch on the lights) is a very simple device that achieves its functionality ('on' and 'off') by cooperative working of different parts (such as connectors, springs, casing etc.) assembled together. However, if we use a piece of doped silicon (thereby 'embedding' the functionality), we can achieve these functionalities ('on' and 'off') by applying a suitable bias voltage. Now if we cut the 'physical switch' into pieces, it will lose its functionalities because of broken parts. However, if we break down the silicon into pieces, each piece will still retain its functionalities (by virtue of being a semiconductor). But the question is: why would anyone be interested in breaking it into pieces? The answer lies in the concept of 'miniaturization' thereby leading to enhanced efficiencies. While there are 'practical limits' for the miniaturization of the 'physical switch', the 'practical limit' for miniaturization of semiconductor devices are many orders of magnitude smaller (we are right now pretty close to this limit: with 5nm fabrication technology commercially introduced by Samsung and TSMC in 2019, our transistors are less than 25 atoms wide). However, this concept of embedding the device functionality enabled us the sustained exponential growth of technology for nearly a century, which is captured in the public parleys in the name of "The Moore's law". This enabling technological platform provides us the perfect launch pad, where machine learning (ML) could be possible.

Before the industrial era, people hardly believed that a machine can beat the muscle power of a human let alone some of the strongest and fastest animals on the planet. That time, 'machines' were just mere 'tools' and lacked raw power. The industrial revolution, particularly the discovery of steam engines, broke that myth. However, humans were still largely content with the faith that they are, and will ever remain, far superior to machines in their intellectual capabilities. The first convincing proof that broke this myth came in 1997 when 'Deep Blue', the IBM supercomputer, defeated the then world chess champion, Garry Kasparov. However, until recently, such amount of huge raw computational power was available to only select entities like giant corporates, defence contractors or to some big national labs. Machine learning would never have been possible like today, if supercomputer-like processing power were not made available to the common people at an unthinkable low price by the rapid advancement of Graphics Processing Units (GPUs) towards the end of last century. Now even a common person has access to both the machine and tools (algorithms) that can beat human-like performance, virtually encompassing each and every area concerning human intelligence, and this is popularly termed as 'Artificial Intelligence' (AI).

## 2 Nature Inspired Engineering: The Birth of AI and Machine Learning

Understanding the working of the human brain has been one of the most challenging and popular scientific and philosophical questions. Complex neurological networks work in perfect tandem to involuntarily control even the minutest bodily functions. Apart from working of the human body, remarkable cognition and continuous learning by our nervous system is what makes us humans an indomitable force in nature. With the aim to seamlessly blend technology in our daily lives, huge efforts are being made world over to develop artificial intelligence for computers which will help and accelerate our daily human activities. Examples include songs, movie recommendations on your search engine, targeted advertisements, and self-driving cars to machine learning driven drug discovery packages that have revolutionised the pharmaceutical industry. Thus, at a time when AI and machine learning is pervading almost all aspects of our lives, it is imperative that the materials research community also reap some of its benefits. After all, as we argued in the previous section, it is the development of materials technology that steers human civilization.

The way scientific enquiry has been conducted in the modern age can divided into four stages known as the paradigms of science as shown in Fig. 2. The first paradigm of science was mostly based on empirical observations from experiments and natural phenomena. Simple experiments like snuffing out of a candle in absence of air ultimately led to the discovery of oxygen as a life-supporting gas. Simultaneously, experimental observations triggered development of many theoretical postulates some of which are now considered as universal laws of physics like the three laws of thermodynamics, the uncertainty principle etc. These formed part of the second paradigm of science. Thereafter, with the advent of powerful calculator and computers in the last few decades, it became possible to almost exactly solve complex equations describing atomic properties like the Schrödinger's equation in density functional theory (DFT) and classical Newton's equation in molecular dynamics (MD). Thus, theory and experiments on materials over the centuries and decades have generated a large amount of data and continue to do so. The fourth paradigm in material science also termed as Materials 4.0 [1] encompasses big data driven representation and analysis thanks to the availability of high throughput computing resources. Consolidating all data into meaningful information for prediction of material behaviour and novel material discovery is one of the ultimate goals of applying general machine learning algorithms in materials research.

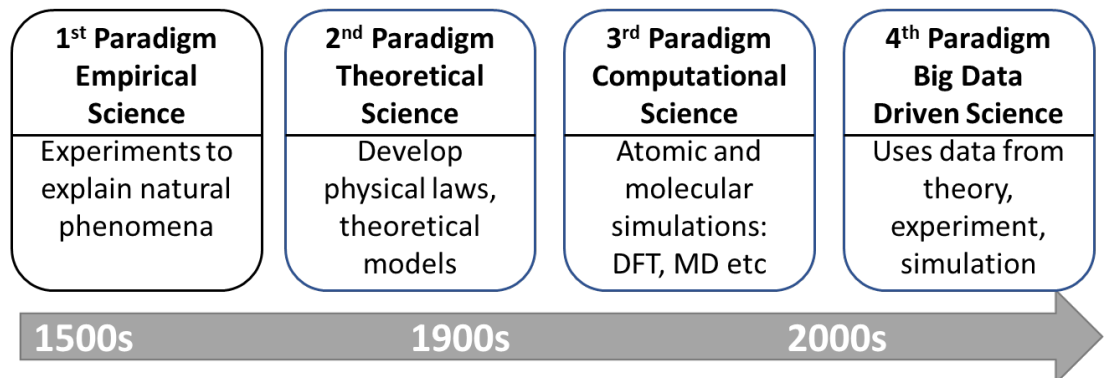

**Fig. 2.** The four paradigms of science through the ages. Adapted from [2].

Machine learning is a subcategory of artificial intelligence (Fig. 3). It is essentially a set of algorithms which can learn from historical data and trends and predict outcomes for further decision making most often without really knowing the underlying physics. These set of algorithms enable systems to improve from experience and keep "learning" without explicit reprogramming.

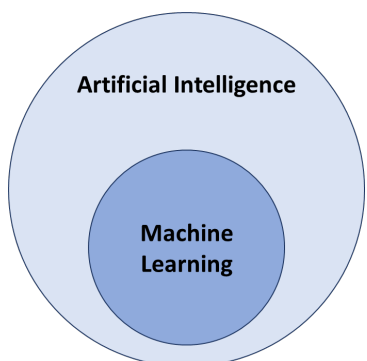

**Fig. 3.** Machine learning techniques as a subset of artificial intelligence regimes.

Machine learning can be broadly classified into 3 major categories:

1. Supervised Learning – Learning from examples, previous data, and historical responses. Like learning from a teacher. Usually implemented where a large amount of test data is available for teaching the code. It involves estimating a mapping function, say *f(x)* which takes input, *x*. The fidelity of this mapping function will determine the accuracy of the output predicted. A good approximation of the mapping function is arrived at only after rigorous training with available input data and their known outcomes. There are conceptually two primary types of supervised learning – classification and regression.

   a. Classification – where the output is expected to be categorised (for e.g.- feasibility of a chemical reaction: yes or no, or to predict if a material is metallic or non-metallic)
   b. Regression – where the predicted output is continuous (for e.g.- predict adsorption energies, atomization energies, band gaps)

2. Unsupervised Learning – It can be used to determine new data pattern and distribution. For example, to cluster similar data into one group. These clusters can then serve as input to machine learning algorithms to learn more about such data. Unsupervised learning is used where not much labelled data (data for which the correct outputs are known a priori) is available.  For example, Principal Component Analysis (PCA), t-distributed Stochastic Neighbour Embedding (t-SNE), community detection etc. are based on unsupervised learning.

3. Semi-supervised Learning – This is the case when labels are available for *few* test data (correct answers are known for *few* cases), but not for the majority. In practice, this is the most prevalent case. However, this is the most challenging area. Most often, it is hard to find a general computational framework that is applicable in other systems as well. Needless to mention, the widest research gap exists here, and it deserves serious attention.

Akin to bioinformatics and genomics, materials informatics aims to consolidate different methodologies of data science and internet technologies to help in materials design and engineering [3]. High-throughput computational techniques are being used to predict material properties and aid in the discovery of photovoltaics [4], thermoelectric materials [5], organic polymer dielectric materials [6], novel topological materials [7], organic LED materials [8] etc. These techniques combine superlative computational power with theoretical constructs like DFT, MD, machine learning assisted screening to help in the discovery of novel materials for myriad applications. Traditional machine learning models applicable to image and signal processing have immense data sets available for training. However, the available databases involving material science problems are tiny in comparison. Machine learning applied to materials can help bridge the gap between experimental and theoretical materials design, synthesis, characterization and modelling [4]. For example, machine learning for chemical synthesis can predict feasible reaction pathways and stability of the product. However, this process is made highly complex by the scarcity of training data and possibility of huge number of feasible solutions. Thus, machine learning on materials presents a unique and unprecedented challenge of working with low volume and widely varied data. This challenge has provided the impetus to improve conventional machine learning models by introducing novel methods and modifications such that they can be applicable to materials. The entire process of implementing machine learning can be broadly divided into three major steps:

1. Data Collection and Representation
2. Model Selection and Validation
3. Model Optimization

# 3 Data Collection and Representation

Nature had the access of only 92 elements. But with only these finite verities it could produce unthinkable varieties of materials. It is even more perplexing to imagine, how using mostly a handful of elements, namely carbon, oxygen, hydrogen and nitrogen it could build extremely huge sets of materials that we name as organic compounds. In the process of evolution of complex living species, it apparently seems that, a nearly perfect material has been synthesized for any particular job. It is the goal of materials engineer to find/develop such capability to meet the ever-growing needs of our civilization. Very obviously, in its simplest form, it essentially boils down to an optimization problem. However, except for very few and simple cases, it is a problem of enormous complexity, which in mathematical terminologies, is called an NP hard problem. In a layman's term, and sufficient for the present discussion, a problem will be called 'NP hard' if an exact solution can NOT be found in finite time using finite resources (mostly referring to computational power). It is actually in this paradigm, that machine learning thrives by providing an acceptable solution within very limited time and resources. It is needless to emphasize the requirement of access to very high quality and precise data for such machine learning models to be successful. Unfortunately, there are not many such resources for materials, despite it being one of the oldest sciences. In the following, we list some of the most important and reliable materials related experimental and theoretical data that are available in open source platforms.

## 3.1 Materials Databases

| Database | |
|---|---|
|  |  |
| ICSD (Inorganic Crystal Structure Database) | 203,830 crystal structure information (https://icsd.fiz-karlsruhe.de) |
| JARVIS (Joint Automated Repository for Various Integrated Simulations) | Database of DFT, MD and ML based calculations (https://jarvis.nist.gov/) |
| Crystallographic Open Database (COD) | 411,160 crystallographic data of organic, inorganic, metal-organics compounds and minerals (http://www.crystallography.net/cod/) |
| GDB-17 | A chemical database with 166 billion small organic molecular structures [9]. |
| QM9 database | Molecular quantum properties for the ~134k smallest organic molecules containing up to 9 heavy atoms (C, O, N, or F; not counting H) in the GDB-17 universe. |
| OQMD (Open Quantum Materials Database) | 0.5 million entries of thermodynamic and structural properties calculated from DFT. (http://oqmd.org/) |
| HOPV15 (Harvard Organic Photovoltaic Dataset) | Database of solar photovoltaic materials under HCEP (Harvard Clean Energy Project) [10]. |
| MP (Materials Project) | Web-based database of computational data under the Materials Genome Initiative (https://materialsproject.org/) |
| NOMAD (Novel Materials Discovery) | Repository of computational data on materials; more than 30 million total energy calculations (https://repository.nomad-coe.eu/) |
| Citrination Platform | Open materials science data and AI platform (https://citrination.com) |

## 3.2 Data Representation

The process of converting raw data into something more suitable for an algorithm is called featurization or feature engineering [4]. Feature space includes data represented in a way meaningful for a machine learning algorithm. Materials can be represented in widely two forms – elemental or structural. Elemental descriptors use the intrinsic properties of the elements, like atomic number, atomization energy, electronegativity, atomic charge etc. Structural descriptors incorporate atomic neighbourhood information. For example – radial distribution functions, Voronoi tessellations [4], Fourier series, graph convolution networks, bond order parameters etc. [11]. Some of the desired features of descriptors are uniqueness, invariance to translation, rotation, symmetry operations, continuous in the feature space, global, and non-degenerate [12, 13]. In fact, sometimes a combination of both structural and elemental descriptors is essential for better representation. Thus, the choice of molecular descriptors is critical to the overall performance of a machine learning algorithm aiming to solve material-based problems. Here we describe a few commonly used material descriptors.

### 3.2.1 Adjacency Matrix

This matrix is very useful for particles with no charge. This is typically a graph representation of the interaction between atoms as shown in Fig. 4. For example, if we are interested only in the bonds between atoms, then the *(i,j)* entry of the matrix will be *1* if a bond exists between atom *i* and *j*, and zero otherwise. The entries in the adjacency matrix need not be binary. One can use bond strength instead of using *1*, thereby allowing all positive real numbers. Typically such matrices are symmetric, because of the symmetry in the interaction between atoms *(i,j)* and *(j,i)*. However, there is no compulsion for this: One can draw a network of atoms that are directed, therefore the adjacency matrix will cease to be symmetric. A specific adaptation of the adjacency matrix, where the chemistry of the atomic species could be incorporated, leads to Coulomb matrix, discussed next.

### 3.2.2 Coulomb Matrices and Bag of Bonds

Coulomb matrices are one of the most popular descriptors which combines both structural and charge information. A typical Coulomb matrix for a molecule is formed by creating a square matrix, *M* with size same as that of the number of atoms in the molecule (Fig. 4). Its form is as given in Eq. (1) where Z are the nuclear charges of the atoms and R are the atomic coordinates [14]. Coulomb matrices have the advantage of being invariant to molecular rotations or symmetry operations. However, a simple Coulomb matrix notation for the same molecule might be different in case the ordering of atoms is changed. This drawback has been overcome by sorting the matrix elements according to eigen values or atoms. Another novel approach is to consider not one but multiple random Coulomb matrix notations for the same molecule as a part of its descriptor set [15].

$$M_{ij} = \begin{cases} 0.5 Z_{ij}^{2.4} & \text{for } i = j \\ \frac{Z_i Z_j}{|R_i - R_j|} & \text{for } i \neq j \end{cases} \qquad \text{Eq. (1)}$$

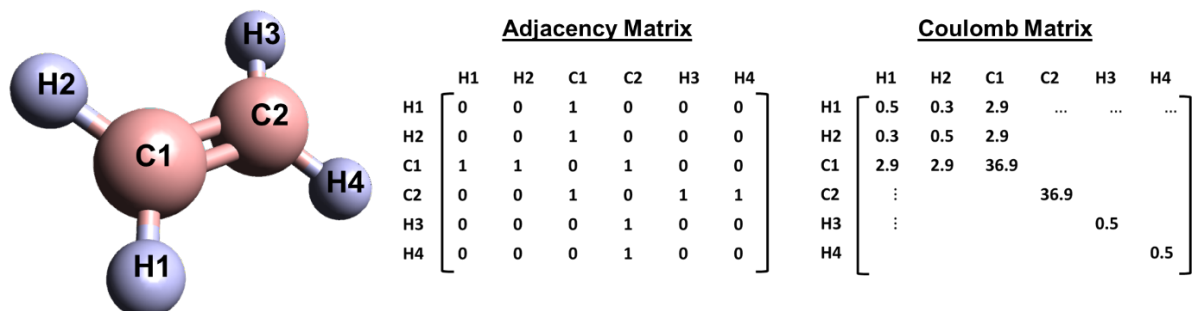

**Fig. 4** Adjacency and Coulomb matrix representation of $C_2H_4$ molecule. Adapted from [15].

Bag of Bonds (BOB) – It is inspired from the bag of words style of representation of data used in natural language processing and was first proposed by Hansen et al for molecules [16]. Like bag of words, each bag denotes a particular type of bond between atoms (such as C-H or H-H) and contains all such types of bonds in the input data set. Coulomb matrix elements can thus be segregated in this scheme and then resorted within each "bag" according to their magnitudes. The bonds, angles, machine learning (BAML) representation is a many-body extension of BOB similar to force fields. They have been used to predict various molecular properties like atomization energies, polarizability, HOMO/LUMO gaps etc. [17].

Other methods of atomic neighbourhood representation are Behler's and Parinello's symmetry functions [18], Bartok's bispectrum method [19] and SOAP (Smooth Overlap of Atomic Positions) based on higher order bond-order parameters [20]. SOAP was found to be a good descriptor for the prediction of hydrogen adsorption energies of hydrogen storage materials [13].

### 3.2.3 Molecular Fingerprinting

It is a popular way of representing a molecule as a set of binary digits indicating presence or absence of certain substructures. It has been widely used in fields from genetics to cheminformatics. Extended connectivity fingerprints (ECFPn) is one such fingerprinting technique. It depicts molecules as a set of subgraphs with a diameter of n.

### 3.2.4 Radial Distribution Functions

Classical pair radial distribution function (RDF) of a crystal calculates the distribution of pair-wise atomic distances and is in itself an excellent descriptor of crystal structure [21]. Fourier series of radial distribution functions, *FR* is an intermolecular distance based descriptor for the chemical compound space (CCS) which satisfies most desired requirements of a descriptor and has been shown to be a good predictor of the potential energy surfaces of molecules [12]. Molecular atomic radial angular distribution (MARAD) is another atomic radial distribution function (RDF) based representation [22].

### 3.2.5 Voronoi Tessellations

Voronoi tessellation is a geometric operation for partitioning Euclidean space. We explore this with help of an example. Suppose we have added *n* nucleating sites randomly in a melt which is uniformly cooling. We assume isotropic growth (that is, all the crystal directions grow at the same rate). Once all the melts cools and from n grains, the resulting grain boundaries will represent a Voronoi diagram. Formally, any point in a Voronoi cell is closer to that centre than any other centre. While in many problems, we use regular grids (like square or cubic grid points), for an arbitrary geometry this might not be optimal. Voronoi tessellation, or its geometric duel, Delaunay triangulation are the most popular methods when uniform grids are not desirable. Voronoi decomposition has been applied to calculate accessible surface areas, void dimensions in porous crystalline materials [23] which can help in establishing Qualitative Structure Property Relationship (QSPR).

### 3.2.6 Principle Component Analysis (PCA)

PCA is a very powerful tool for dimensionality reduction and therefore used in unsupervised learning. It is based on the calculation of the eigenvalues and corresponding eigenvectors. This method has a very sound mathematical foundation, which is based upon linear algebra. Therefore, the interpretation of the results is straightforward. The main shortcoming is the underlying assumption of linearity. Since most material science problems are non-linear, therefore it is very rarely used in advanced applications for machine learning. PCA has been used to filter out irrelevant densities in order to be used by kernel ridge regression models to predict the kinetic energy of a model of 1 dimensional non-interacting fermions [24].

### 3.2.7 T-Distributed Stochastic Neighbour Embedding (t-SNE)

Generally, materials data have multiple features. Implied is the fact that such data, more often than not, resides in a high dimensional space. Visualization, presentation, perception and even at times for analysis, dimensionality reduction of the data is a prime requirement. T-distributed stochastic neighbour embedding (t-SNE), is one of the most powerful techniques for visualization of high-dimensional data onto a lower dimension. This is, in essence, a non-linear dimensionality reduction technique based upon the 'Student's t-distribution'. This is very often used to represent unbiasedness of data and correlation of descriptors at a lower complexity level (reduced dimension). t-SNE has been used to cluster 267 hybrid bilayer 2 dimensional materials for predicting their band gap and interlayer distances [25].

### 3.2.8 Molecular Graph Representation

Molecular graphs can serve as input to neural networks. Standardizing all the features is important so that all the property have zero mean and unit standard deviation. Some attributes used in the MEGNet (MatErials Graph Network) framework developed for predicting molecular and crystal properties are given in Table 1 [26]. Molml is a python library to create different molecular representations for use in machine learning models.

**Table 1:** Some features in a molecular graph representation for input to neural networks [26].

| Feature | Description |
| --- | --- |
| Atom Attributes | Atom type, atomic number, electronic charge, hybridization |
| Bond Attributes | Bond type (single or double), bond existence, bond order, distance between atoms |
| State Attributes | Temperature and pressure of the system, average atomic weight, average number of bonds |

### 3.2.9 Community Detection

Originally proposed by social scientists primarily to investigate the social investigation, community detection, later became one of the most powerful tools for unsupervised learning when improved and adapted by the computational fraternity. However, its use did not penetrate deeply in materials science. Pioneering works by Aharonov [27] to study some toy model of glass structure produced exciting materials model system for trying out machine learning. While many variants of this algorithm (community detection) exist, one of the most popular methods is inspired by the mathematical framework originally proposed to model the Hamiltonian (total energy) of a spin-glass system [28]. Huge improvement in this scheme was achieved when Nussinov and Ronhovde [29, 30] proposed implementation of a new parameter (they called it $\gamma$, the resolution) so the algorithm could scan the entire length scale (from smallest length to the largest length that the system incorporates) in order to mine the data to find the patterns of interest. They further showed that, if this community detection algorithm is coupled with metrics for comparing the information content (like the normalized mutual information, or NMI; variance of information, or VI; entropy of information etc.) then the algorithm is capable to auto adjust the relevant length scale of interest and can identify hierarchical structures. This idea was successfully employed for analysing the structure of $Al_{88}Y_7Fe_5$ metallic glass [29]. Further improvement of this scheme was proposed by Raj et al. [31] for better adaptation of the community detection scheme for materials systems. The objective of the community detection scheme is to find the best partition of the data. However, one needs a metric to quantify the 'quality of partition'. Modularity, originally proposed by Newman [32, 33], is arguably the most popular method for quantifying this 'quality of partition'. While it is very well suited for the systems that are not embedded in Euclidean space, for example analysing the words in a book; it performs sub-optimally for spatially embedded networks. The modularity function proposed by Raj et al based on the ideas of the Hamiltonians of the spin glass system. Using 2D and 3D systems of granular packing, it was established that this method produces far superior results than the competing methods.

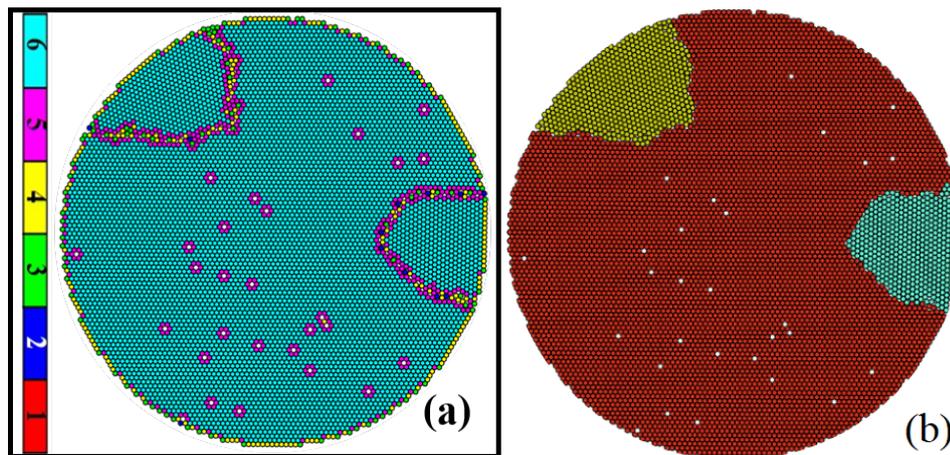

Fig 5: The (a) input and (b) output to the unsupervised machine learning algorithm [31]. Permission taken from authors.

A good descriptor is quantified by the kernel function *k(q,q')* which gives the similarity between descriptor sets, *q*. More similarity indicates better descriptor. Before designing a model, it is also important to assess the correlation between the various input descriptors and the output variables to be predicted. A simple linear correlation can be established by calculating the Pearson correlation coefficients (PCC). The Pearson correlation is calculated by dividing the covariance of two variables by the product of their standard deviations. A PCC value of 1 indicates complete linear correlation between input and output variables, whereas a PCC value of 0 indicates no linear correlation at all. For example, electron affinity is a necessary variable to predict the redox potentials of organic compounds, whereas, a quantity like the count of aromatic rings has no correlation at all with redox potentials [34].

## 4 Model Selection and Validation

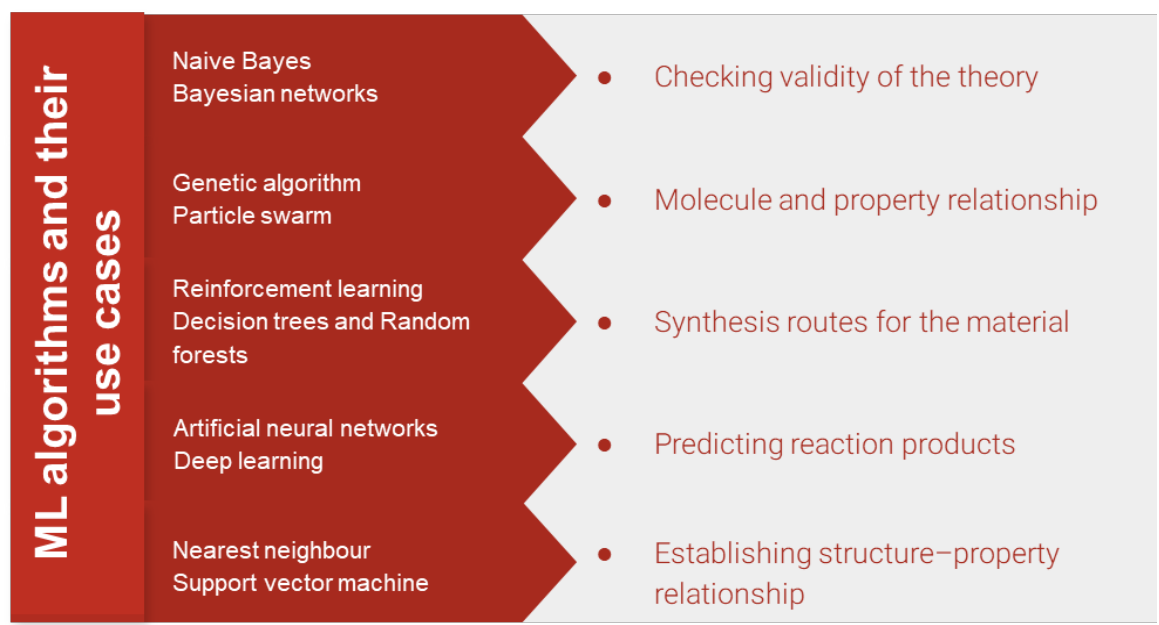

Fig. 6. Choice of machine learning models depicting few examples [4].

There are numerous machine learning models or algorithms available. Some are as simplistic as fitting to a linear regression curve and some extremely complex like neural networks. As discussed earlier, materials world presents a unique challenge to machine learning algorithm developers because of the scarcity in data available compared to other fields, absence of a uniform protocol for recording experimental and theoretical data, presence of unverified and spurious data as well. Thus, conventional models have not been very successful. Here we discuss some algorithms used on materials data. Although, machine learning for materials is a pretty recent avenue of science, it is at a nascent stage where many researchers are designing new algorithms. As argued earlier, for application of machine learning in materials, not algorithms but availability of large datasets of high quality is the true bottleneck right now. With the advent of highly accurate density functional theory regimes, in near future there might not be dearth of theoretical data. However, performing full DFT calculation on a large system of different atom types is still a computationally costly affair. Machine learning can be used to replace first-principles calculations and reducing computational cost and user time. The total energy of a molecule as a function of electronic density, *n* is given by the Kohn-Sham equation

$$E(n) = T_s(n) + U_H(n) + V(n) + E_{xc}(n)$$

Where, $T_s$ is the kinetic energy of non-interacting electrons, $U_H$ is the Hartree potential energy, V is the external potential and $E_{xc}$ is the exchange-correlation term appearing due to energy approximations in the Kohn-Sham approach. An accurate total energy calculation is computationally intensive. With machine learning approaches, one can hope to predict the outcomes of a DFT simulation without actually performing it. Such a bypassing of Kohn-Sham equations in a DFT calculation by learning the energy functional was achieved by kernel ridge regression algorithm [35]. Another approach is to use abundantly available experimental data (like XRD data) as input fed to machine learning algorithms and establish structure-property relationships.

One of the first sciences where machine learning models were applied was the chemical and pharmaceutical sciences. Drug discovery is one of the first avenues to have explored the utility of machine learning in new drug molecule identification and protein engineering. These models aim to map drug molecular structure to target activities. Quantitative Structure-Activity Relationship (QSAR) modelling is a popular model used to assess the cost, time and optimized pharmacodynamics and pharmacokinetics properties [36]. Various QSAR approaches like linear discriminant analysis (LDA), support vector machines (SVM), decision trees (DT), random forest (RF), k-nearest neighbour (KNN), and artificial neural network (ANN) have been employed for drug discovery. Such models have successfully predicted anti-fungal activity of drugs from 3-D topographic descriptors, classified anti-HIV peptides, bioactivity classification, toxicity modelling and drug target identification. However, simple QSAR models are unable to tackle large amounts of data as well as varied training sets. Hence, alternative deep learning techniques and combinatorial QSAR techniques which amalgamate more than one QSAR models are being investigated for better predictions in drug discovery [36]. DELPHOS and MoDeSuS are state-of-the-art QSAR based models which have been used to predict BOD of chemical compounds [37]. Here we discuss some widely used ML algorithms and their modifications made suitable for materials.

## 4.1 Regressors

It includes linear models (Bayesian ridge regression (BR) and linear regression with elastic net regularization (EN)), random forest (RF), kernel ridge regression (KRR), and two types of neural networks, graph convolutions (GC) and gated graph networks (GG). There are specific combinations of regressors and representations for different features (properties). The results suggest that ML models could be more accurate than hybrid DFT if explicitly electron correlated quantum (or experimental) data were available [22].

### 4.1.1 Kernel Regression

Kernel-based learning methods are one of the most widely used algorithms. Kernels measure the similarity between two data sets as input. The output is a linear combination of kernel functions for the data set. We have a set of data $\{P_i\}$ described as a function of $\{x_i\}$ descriptors. The general form of kernel regression is

$$P_i = P(x_i) = \sum_{k=1}^{N} c_k K(x_i, x_k)$$

| Linear regression | $K(x_i, x_k) = x_i \cdot x_k$ |
| | $P(x_i) = x_i . c$ |
| Polynomial kernel regression | $K(x_i, x_k) = (x_i \cdot x_k + c)^d$ |
| Gaussian kernel regression | $K(x_i, x_k) = exp\left(\dfrac{-\sum_j (x_i - x_k)^2}{2\sigma_j^2}\right)$ |

Gaussian fit is a very popular data fitting method where kernel regression uses a sum of multiple Gaussian curves to fit data in machine learning. Kernel ridge regression models are more flexible but less efficient than linear regression models. These models have been used to predict potential energy surfaces [38, 39], electronic density of states [21], formation energies [40] etc.

## 4.2 Neural Networks

Artificial neural networks (ANNs) are modelled after the brain. It is an algorithm which takes input (features) and gives the output (predictions). In between there is at least one "hidden layer". The hidden layers consist of some activation functions (like the sigmoid function) which weighs the input and generates an output for the next layer. A deep neural network is one with many such hidden layers. Neural networks are one of the best algorithms for machine learning but need a large amount of data and are also computationally intensive. But it has been shown to work with low cost and good accuracy for smaller data sets to predict redox potentials of possible battery electrode materials by using electronic and structural descriptors as input with an error of less than 4% [34]. Neural Designer is a machine learning platform where we can design and optimize custom neural networks.

One of the advantages of neural networks is that complex forms of data representation is no longer a separate requirement. The model "learns" the best featurization from primitive information about molecules, like interatomic distance and charge, on-the-go. It is able to capture underlying irregularities. Even though neural networks work best with large amounts of data, newer modifications have made it possible for neural networks to work with small amounts of data as well. For example, ElemNet is a deep neural network which can work better than conventional machine learning models on a dataset of just 4000 compounds [41]. Deep tensor neural network (DTNN) approach can learn the potential energies of organic molecules in a molecular dynamics trajectory with an accuracy of 1 kcal mol$^{-1}$ from a data set of 300,000 to 900,000 timesteps [42]. Grossman's model of a generalized crystal graph convolutional neural network (CGCNN) has accurately predicted 8 different properties of perovskite crystals after training with 10,000 data points only. Remarkably, the accuracy is as high or even more than conventional DFT predictions as compared to experimental results [43]. Another novel approach is coupling genetic algorithms with powerful neural networks by employing ANNs to either evaluate the genetic algorithm's fitness function or introduce a bias towards certain data [44]. This strategy successfully circumvents the problem of limited data availability in materials science for efficient neural network training.

## 4.3 Transfer Learning

Until now, we have discussed some widely used ML algorithms. However, as mentioned earlier, insufficient data is a critical bottleneck in applying these algorithms to material science. Transfer learning is a novel technique which deals with this major challenge by transferring information between learning tasks. Four architectures for transfer have been tested by Hutchinson et al. to predict band gaps and colour of crystalline compounds at 300K [45]. One of these schemes called the difference architecture, where the difference between responses of different models is learned by another model. By learning the difference between experimental and computational band gaps, this scheme was able to predict the band gaps with high fidelity using just one fourth of the experimental data as used by a conventional baseline model [45].

## 4.4 Natural Language Processing for Materials Literature

Natural language processing (NLP) is a machine learning technique which processes human language. It is also used to potentially create human language from learning language data as inputs. Popular examples are how search engines like Google returns results relevant to the keywords inserted, Google's language translation application, mail spam detectors, auto-correct in our smartphone keyboards, speech recognition etc. Thus, NLP can be expected to process textual data of any kind including materials-based literature. In an interesting study conducted by Kim et. al., NLP techniques were used to scan through more than 12,000 manuscripts to learn the critical parameters required for hydrothermal synthesis of titania nanotubes [46].

## 4.5 Machine Learning Toolkits

With the possibility of using a variety of machine learning algorithms on materials data, it is important that such code be made available in a tool framework for wider usage. One such open source machine language framework is the *Amp* (Atomistic Machine-learning Package) [47]. *Amp* implements fast machine learning to predict potential energy surfaces by interpolating from a training data set of potentials. It is currently integrable with the most popular DFT codes for generating the training potentials, VASP, GPAW, Dacapo, Quantum ESPRESSO under the Atomistic Simulation Environment. Table 2 lists some other general-purpose packages and libraries in R and Python languages that can be used for materials modelling using machine learning and a few other tools and libraries exclusive to materials.

Table 2. Few general-purpose and materials specific machine learning frameworks and tools in R and Python.

| Package Name | Language |
|---|---|
| Caret | Functions based on R for classification and regression models (https://github.com/topepo/caret) |
| H2O.ai | Java-based data modelling which is easily accessible by Python and R based modules. (https://www.h2o.ai) |
| Scikit-learn | SciPy Python library. (https://scikit-learn.org/) |
| Keras | Open source neural network Python library. (http://keras.io/) |
| Tensorflow | Deep learning toolkit in Python. (https://www.tensorflow.org/) |

| pyTorch | Python library for various ML tools. (https://pytorch.org/) |
|---|---|
| Machine learning frameworks for materials | |
| AMP (Atomistic Machine-learning Package) | Machine learning for atomistic calculations (https://amp.readthedocs.io/en/latest/) |
| MAterials Simulation Toolkit for Machine Learning (MAST-ML) | Open-source Python package (https://github.com/uw-cmg/MAST-ML) |
| Materials-Agnostic Platform for Informatics and Exploration (MAGPIE) | Java based library of algorithms (https://bitbucket.org/wolverton/magpie) |
| Molml | Python library for representation of molecular data. (https://pypi.org/project/molml/) |
| COMBO (COMmon Bayesian Optimization Library) | Python library for ML techniques (https://github.com/tsudalab/combo) |
| MatMiner | Python library for mining materials data. (https://github.com/hackingmaterials/matminer) |
| TensorMol | Python package for neural network based molecular simulations. (https://github.com/jparkhill/TensorMol) |

Rule-based approaches like Boltzmann machines have been somewhat successful. Others like Organic Chemical Simulation of Synthesis programme (OCSS), sequence-to-sequence approaches (organic chemistry and linguistics) and one-shot learning.

Regressors can roughly be ordered by performance, independent of property and representation. In general, neural networks like graph convolution and gated graph neural networks in conjunction with appropriate methods of data representation have higher performance in predicting certain ground-state properties of organic molecules [22, 34].

## 5  Model Optimization and Quality Assessment

Validation or test data sets are used to estimate the performance of the machine learning model created. Errors in prediction of the output from a known set of inputs in supervised learning are known as prediction errors, which may be of 2 types –

- Bias errors – It depicts how much the predicted output differs from the actual known output of a sample test data. A high bias error suggests that the mapping function is too approximate, probably formulated for developing a faster and simpler algorithm. A low bias on the other hand, suggests fewer approximations were made in the mapping function leading to higher, sometimes unnecessary computational cost.
- Variance errors – It quantifies the difference between algorithm outputs when different training data sets are used. Ideally, a mapping function should be immune to changing training data sets. However, few ML algorithms like decision trees, support vector machines and k-nearest neighbour have been known to suffer from high variance issues.

Both errors if not balanced properly gives rise to crippling problems like overfitting and underfitting. Underfitting occurs when the mapping function misses out on important details of the data. It is indicative of high bias and low variance. On the other hand, overfitting occurs when the model has low bias and high variance and ends up capturing not only the important details but also a lot of unnecessary noise.

Thus, the model chosen should be optimized rigorously by arriving at a suitable trade-off between bias and variance to avoid the pitfalls of over and underfitting. The various metrics for model evaluation are:
- Loss function
- Mean absolute error
- Learning rate
- Mean absolute relative error
- Coefficient of determination $R^2$
- Receiver operating characteristic curve (ROC) – area under ROC; close to 100% area implies better classification model.

Newton's optimization method calculates Hessian matrix of $2^{nd}$ order derivative of the loss function. Quasi-Newton optimization method calculates the inverse Hessian matrix using the first order derivative of loss function and does not need to determine the exact value of the Hessian matrix.

# 6 Opportunities of ML for 2D Materials

## 6.1 Why do we need ML for 2D materials research?

Recent advances in atomically thin two dimensional (2D) materials have led to a variety of promising future technologies for post-CMOS nanoelectronics, energy, photonics and optoelectronics [48-52]. 2D materials are comprised of layered van der Waals (vdW) solids such as atomically flat graphene, h-BN, phosphorene, SiC, $Si_2BN$, Transition Metal Dichalcogenides (TMDs, $MX_2$ (X-M-X layer) where M = Ti, Zr, Hf, V, Nb, Ta, Mo, W, and X = S, Se, Te). Other 2D materials such as MXenes (e.g., $Ti_3C_2$) are gaining enormous interest for its applicability in various disciplines [53,54]. Though 2D materials have a wide range of applications, the bottleneck of everything is their synthesis with air-stability and less/no defects. Moreover, in spite of the recent advances in the synthesis and characterization of 2D materials, serious issues related to materials quality are a roadblock to advancing their science and applications. The presence of defects significantly alters various properties of 2D materials [55]. In addition, during growth, oxygen atoms get adsorbed (oxidation) on the surface of the 2D materials [56]. The oxidation induces deterioration of 2D materials [57]. The experimental approaches to investigate these issues are always associated with the high cost and human resources. Hence, it is necessary to have the computational prediction to facilitate the efficient experimental design for real-life applications. Machine learning (ML)-based prediction can be a viable and optimal approach to accurately predict the possible synthesizability of 2D materials with controlled defects. Similarly, ML can predict which 2D materials among a wide range of options can be the best choice for a particular application. For example, 2D materials have enormous applications in energy storage [58]. However, it's costly and inefficient to predict experimentally the best 2D materials and their surface topology for a particular type of battery applications (e.g., lithium-ion battery). ML can predict efficiently the most effective 2D materials for a particular type of battery. In the next two sections, we will briefly discuss the application of ML for prediction of properties of 2D materials and their application in sustainable energy storage.

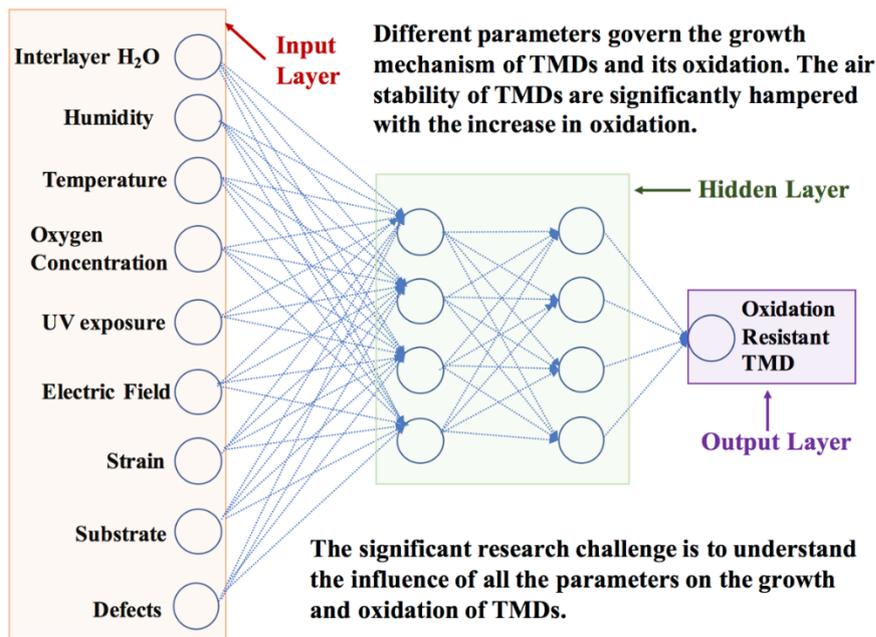

**Fig. 7**. Machine learning problem to predict the most air-stable Transition Metal Dichalcogenide (TMD). The approach is similar for other 2D materials.

## 6.2 Machine learning to predict the properties and synthesizability of 2D materials

The input layers of Fig. 7 show the various parameters that typically control the growth mechanism of 2D materials [57]. Considering multiple permutation and combinations of the available 2D materials, and taking into consideration many parameters affecting the growth process, it's a daunting task for experimentalists to explore all possibilities for finding stable 2D materials. However, Fig. 7 shows that the problem can be formulated as an ML problem. Recently, the Positive and Unlabeled (PU) ML algorithm has been implemented for prediction of synthesis of 2D metal carbides and nitrides (MXenes) and their precursors [59]. Using the elemental information and data from high-throughput density functional theory (DFT) calculations, PU learning methods was applied to the MXene family of 2D transition metal carbides, carbonitrides, and nitrides, and their layered precursor MAX phases, and 18 MXene compounds were identified to be highly promising for synthesis. Considering the ML formation in Fig. 7, detailed studies are necessary for the prediction of the synthesizability of other 2D materials such as TMDs.

Besides synthesizability, ML has been implemented to predict the band gap of functionalized MXene [60], where ML models can bypass the band gap underestimation problem of local and semilocal functionals used in DFT calculations, without subsequent correction using the time-consuming GW approach. ML approaches have been integrated for accelerated discovery of TMDs as elemental mercury (Hg°) sensing materials [61]. However, the application of ML for predicting the properties of 2D materials is still at a nascent stage. For example, oxidation is a major problem is producing air-stable TMDs (Fig. 8a) [57]. Some investigations suggest that TMDs, grown on various other 2D substrates (e.g., graphene, hBN) placed or suspended on $SiO_2$, can reduce the oxidation [57]. The presence of defects (Fig. 8b) significantly influences the oxidation behavior. However, there are various types of defects and 2D materials [55]. The ML approaches are necessary to systematically find the best 2D materials as a substrate for the air stability of a particular TMDs. Moreover, ML algorithms can predict the possibility of the presence of a defect type in TMDs, and how does this defect influence the oxidation and other properties of TMDs efficiently.

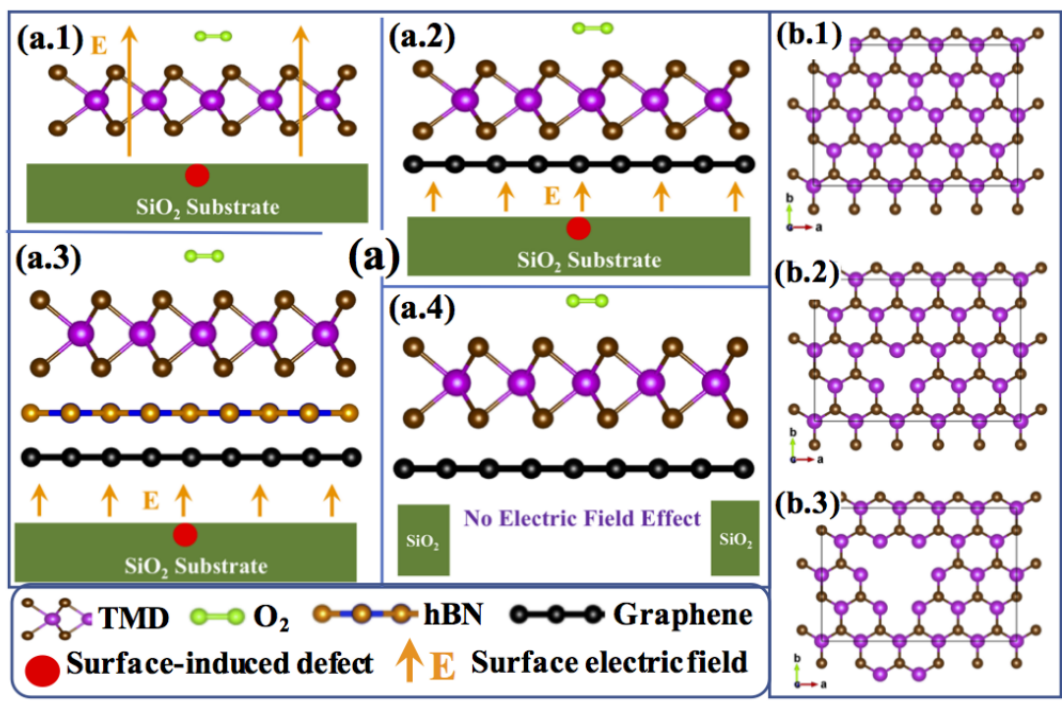

**Fig. 8**: (a) Oxidation of TMD on various substrates such as (a.1) $SiO_2$, (a.2) graphene on $SiO_2$, (a.3) hBN and graphene on $SiO_2$, (a.4) suspended graphene. (b) Various defects in TMDs.

## 6.3 Opportunities of ML for 2D Materials in Energy Storage

ML has been extensively applied for battery research over the last few years. Using the Materials Project database, the holistic computational structural screening of more than 12,000 candidates was performed for solid lithium-ion conductor materials [62]. Artificial neural network (ANN) has been applied to design the lithium-ion batteries (LIBs) [63, 64]. ML is also integrated with DFT (DFT-based ML) for developing molecular electrode materials in LIBs [34]. All these studies on ML for energies are not on 2D materials based systems. However, despite the recent advances, batteries have many problems where 2D materials can provide a viable solution. In Chapter 5.3, a detailed discussion is provided on the application of 2D materials for energy storage. Here, we discuss how ML can be useful for 2D materials-based batteries.

2D materials have enormous applications in all segments of battery architecture. Fig. 9a shows the schematic of an anode. Interface failure is one of the crucial problems in batteries (Fig. 9b) [65]. 2D materials can be used as a coating on the current collector [65]. This arrangement will produce a van der Waals (vdW) slippery surface to reduce the stress at the interface and battery failure. Similarly, 2D materials can replace the traditional polymer binder to avoid the problem of binder fracture (Fig. 9b1) [66]. There is experimental evidence that graphene can be used as a vdW slippery interface for the current collector, and MXene can be used instead of the polymer binder [66]. However, there are over 700 2D materials, and considering different defect types, the number of options is enormous. Particular 2D materials may be suitable for some specific active materials. For example, graphene may be a suitable vdW slippery surface for silicon anode but not for the tin anode. Doing DFT studies on various bulk-2D materials interface (e.g., graphene - silicon) to find the best candidate involves a high computational cost and human resources. ML is necessary for fast and accurate prediction of 2D materials for specific active materials.

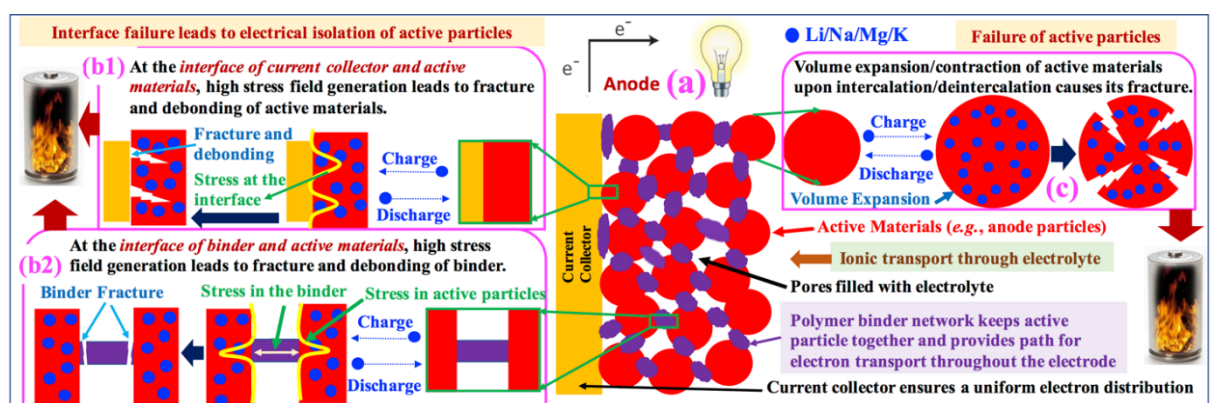

**Fig. 9**: (a) Schematic of a battery. (b) Failure at the interface – *(b1)* interface of current collector and the active materials. *(b2)* interface of binder and active materials. (c) Failure of active materials.

Besides interface failure, the most common failure mechanism in batteries are the fracture of active materials (Fig. 9c) [67]. Because of the volume expansion-contraction of active materials upon intercalation-deintercalation, active particle breaks. 2D materials-based anode materials (Fig. 10a) can accommodate more ions without substantial volume change [67]. Hence, 2D materials-based anode can yield very high capacity, power density, and energy density. Moreover, recent reports show that nanofluid confined 2D materials (Fig. 10b) anode can facilitate in faster diffusion leading to enhanced cycle life [68]. However, with over 700 2D materials, the number of possible heterostructures is over a million. Moreover, if nanoconfined fluid (Fig. 10b) is considered, the possibilities are even higher. In addition, if different defects are considered, the number of possibilities is almost infinite. It's not possible for manual DFT calculations to find the best possible combination of 2D materials and its heterostructures for a particular battery applications. ML approaches can make use of the existing experimental and computational data for the viable prediction of 2D materials-based anode materials for high-capacity, power density, and energy density batteries.

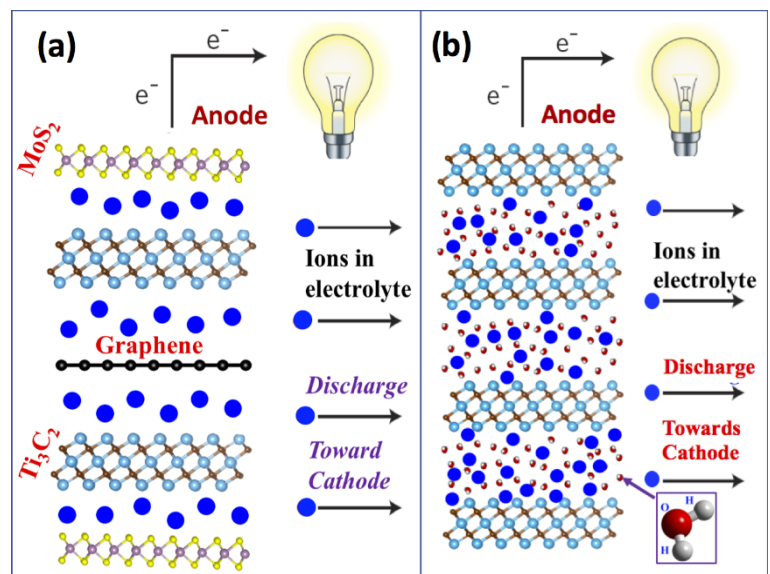

**Fig. 10.** (a) 2D materials and its heterostructures as anode materials for energy storage. (b) Nanofluid confined 2D materials for energy storage.

# 7   References:

[1] Jose, R., & Ramakrishna, S. (2018). Materials 4.0: Materials big data enabled materials discovery. *Applied Materials Today*, *10*, 127–132. doi: 10.1016/j.apmt.2017.12.015
[2] Hey, A. J. G., Tansley, S., & Tolle, K. M. (2009). *The fourth paradigm: data-intensive scientific discovery*. Redmond, WA: Microsoft Research.
[3] Ramakrishna, S., Zhang, T. Y., Lu, W. C., Qian, Q., Low, J. S. C., Yune, J. H. R., ... & Kalidindi, S. R. (2019). Materials informatics. *Journal of Intelligent Manufacturing*, *30*(6), 2307-2326.
[4] Butler, K. T., Davies, D. W., Cartwright, H., Isayev, O., & Walsh, A. (2018). Machine learning for molecular and materials science. *Nature*, *559*(7715), 547.
[5] Gorai, P., Stevanović, V., & Toberer, E. S. (2017). Computationally guided discovery of thermoelectric materials. *Nature Reviews Materials*, *2*(9), 17053.

[6] Kim, C., Pilania, G., & Ramprasad, R. (2016). Machine learning assisted predictions of intrinsic dielectric breakdown strength of ABX3 perovskites. *The Journal of Physical Chemistry C*, *120*(27), 14575-14580.

[7] Autès, G., & Yazyev, O. V. (2018). Discovery of Novel Topological Materials Via High-throughput Computational Search. In *Computational Materials Discovery* (pp. 392-422).

[8] Gómez-Bombarelli, R., Aguilera-Iparraguirre, J., Hirzel, T. D., Duvenaud, D., Maclaurin, D., Blood-Forsythe, M. A., ... & Markopoulos, G. (2016). Design of efficient molecular organic light-emitting diodes by a high-throughput virtual screening and experimental approach. *Nature materials*, *15*(10), 1120.

[9] Ruddigkeit, L., Van Deursen, R., Blum, L. C., & Reymond, J. L. (2012). Enumeration of 166 billion organic small molecules in the chemical universe database GDB-17. *Journal of chemical information and modeling*, *52*(11), 2864-2875

[10] Aspuru-Guzik, A.. (2016, March 1). The Harvard Organic Photovoltaics 2015 (HOPV) dataset: An experiment-theory calibration resource. (Version 4). figshare. doi:10.6084/m9.figshare.1610063.v4

[11] Choudhary, K., DeCost, B., & Tavazza, F. (2018). Machine learning with force-field-inspired descriptors for materials: Fast screening and mapping energy landscape. *Physical Review Materials*, *2*(8), 083801.

[12] Von Lilienfeld, O. A., Ramakrishnan, R., Rupp, M., & Knoll, A. (2015). Fourier series of atomic radial distribution functions: A molecular fingerprint for machine learning models of quantum chemical properties. *International Journal of Quantum Chemistry*, *115*(16), 1084-1093.

[13] Jäger, M. O., Morooka, E. V., Canova, F. F., Himanen, L., & Foster, A. S. (2018). Machine learning hydrogen adsorption on nanoclusters through structural descriptors. *npj Computational Materials*, *4*(1), 37.

[14] Rupp, M., Tkatchenko, A., Müller, K. R., & Von Lilienfeld, O. A. (2012). Fast and accurate modeling of molecular atomization energies with machine learning. *Physical review letters*, *108*(5), 058301.

[15] Hansen, K., Montavon, G., Biegler, F., Fazli, S., Rupp, M., Scheffler, M., ... & Müller, K. R. (2013). Assessment and validation of machine learning methods for predicting molecular atomization energies. *Journal of Chemical Theory and Computation*, *9*(8), 3404-3419.

[16] Hansen, K., Biegler, F., Ramakrishnan, R., Pronobis, W., Von Lilienfeld, O. A., Müller, K. R., & Tkatchenko, A. (2015). Machine learning predictions of molecular properties: Accurate many-body potentials and nonlocality in chemical space. *The journal of physical chemistry letters*, *6*(12), 2326-2331.

[17] Huang, B., & Von Lilienfeld, O. A. (2016). Communication: Understanding molecular representations in machine learning: The role of uniqueness and target similarity.

[18] Behler, J., & Parrinello, M. (2007). Generalized neural-network representation of high-dimensional potential-energy surfaces. *Physical review letters*, *98*(14), 146401.

[19] Bartók, A. P., Payne, M. C., Kondor, R., & Csányi, G. (2010). Gaussian approximation potentials: The accuracy of quantum mechanics, without the electrons. *Physical review letters*, *104*(13), 136403.

[20] Bartók, A. P., Kondor, R., & Csányi, G. (2013). On representing chemical environments. *Physical Review B*, *87*(18), 184115.

[21] Schütt, K. T., Glawe, H., Brockherde, F., Sanna, A., Müller, K. R., & Gross, E. K. U. (2014). How to represent crystal structures for machine learning: Towards fast prediction of electronic properties. *Physical Review B*, *89*(20), 205118.

[22] Faber, F. A., Hutchison, L., Huang, B., Gilmer, J., Schoenholz, S. S., Dahl, G. E., ... & Von Lilienfeld, O. A. (2017). Prediction errors of molecular machine learning models lower than hybrid DFT error. *Journal of chemical theory and computation*, *13*(11), 5255-5264.

[23] Willems, T. F., Rycroft, C. H., Kazi, M., Meza, J. C., & Haranczyk, M. (2012). Algorithms and tools for high-throughput geometry-based analysis of crystalline porous materials. *Microporous and Mesoporous Materials*, *149*(1), 134-141.

[24] Snyder, J. C., Rupp, M., Hansen, K., Müller, K. R., & Burke, K. (2012). Finding density functionals with machine learning. *Physical review letters*, *108*(25), 253002.

[25] Tawfik, S. A., Isayev, O., Stampfl, C., Shapter, J., Winkler, D. A., & Ford, M. J. (2019). Efficient prediction of structural and electronic properties of hybrid 2D materials using complementary DFT and machine learning approaches. *Advanced Theory and Simulations*, *2*(1), 1800128.

[26] Chen, C., Ye, W., Zuo, Y., Zheng, C., & Ong, S. P. (2019). Graph networks as a universal machine learning framework for molecules and crystals. *Chemistry of Materials*, *31*(9), 3564-3572.

[27] Aharonov, E., Bouchbinder, E., Hentschel, H. G. E., Ilyin, V., Makedonska, N., Procaccia, I., & Schupper, N. (2007). Direct identification of the glass transition: Growing length scale and the onset of plasticity. *EPL (Europhysics Letters)*, *77*(5), 56002.

[28] Nussinov, Z., Ronhovde, P., Hu, D., Chakrabarty, S., Sun, B., Mauro, N. A., & Sahu, K. K. (2016). Inference of hidden structures in complex physical systems by multi-scale clustering. In *Information Science for Materials Discovery and Design* (pp. 115-138). Springer, Cham.

[29] Ronhovde, P., Chakrabarty, S., Hu, D., Sahu, M., Sahu, K. K., Kelton, K. F., ... & Nussinov, Z. (2011). Detecting hidden spatial and spatio-temporal structures in glasses and complex physical systems by multiresolution network clustering. *The European Physical Journal E*, *34*(9), 105.

[30] Ronhovde, P., Chakrabarty, S., Hu, D., Sahu, M., Sahu, K. K., Kelton, K. F., ... & Nussinov, Z. (2012). Detection of hidden structures for arbitrary scales in complex physical systems. *Scientific reports*, *2*, 329.

[31] Kishore, R., Gogineni, A. K., Nussinov, Z., & Sahu, K. K. (2019). A nature inspired modularity function for unsupervised learning involving spatially embedded networks. *Scientific reports*, *9*(1), 2631.

[32] Newman, M. E., & Girvan, M. (2004). Finding and evaluating community structure in networks. *Physical review E*, *69*(2), 026113.

[33] Newman, M. E. (2004). Fast algorithm for detecting community structure in networks. *Physical review E*, *69*(6), 066133.

[34] Allam, O., Cho, B. W., Kim, K. C., & Jang, S. S. (2018). Application of DFT-based machine learning for developing molecular electrode materials in Li-ion batteries. *RSC advances*, *8*(69), 39414-39420.

[35] Brockherde, F., Vogt, L., Li, L., Tuckerman, M. E., Burke, K., & Müller, K. R. (2017). Bypassing the Kohn-Sham equations with machine learning. *Nature communications*, *8*(1), 872.

[36] Zhang, L., Tan, J., Han, D., & Zhu, H. (2017). From machine learning to deep learning: progress in machine intelligence for rational drug discovery. *Drug discovery today*, *22*(11), 1680-1685.

[37] Martínez, M. J., Razuc, M., & Ponzoni, I. (2019). MoDeSuS: A Machine Learning Tool for Selection of Molecular Descriptors in QSAR Studies Applied to Molecular Informatics. *BioMed research international*, *2019*.

[38] Hu, D., Xie, Y., Li, X., Li, L., & Lan, Z. (2018). Inclusion of machine learning kernel ridge regression potential energy surfaces in on-the-fly nonadiabatic molecular dynamics simulation. *The journal of physical chemistry letters*, *9*(11), 2725-2732.

[39] Ferré, G., Haut, T., & Barros, K. (2017). Learning molecular energies using localized graph kernels. *The Journal of chemical physics*, *146*(11), 114107.

[40] Faber, F. A., Lindmaa, A., Von Lilienfeld, O. A., & Armiento, R. (2016). Machine Learning Energies of 2 Million Elpasolite (A B C 2 D 6) Crystals. *Physical review letters*, *117*(13), 135502.

[41] Jha, D., Ward, L., Paul, A., Liao, W. K., Choudhary, A., Wolverton, C., & Agrawal, A. (2018). Elemnet: deep learning the chemistry of materials from only elemental composition. *Scientific reports*, *8*(1), 17593.

[42] Schütt, K. T., Arbabzadah, F., Chmiela, S., Müller, K. R., & Tkatchenko, A. (2017). Quantum-chemical insights from deep tensor neural networks. *Nature communications*, *8*, 13890.

[43] Xie, T., & Grossman, J. C. (2018). Crystal graph convolutional neural networks for an accurate and interpretable prediction of material properties. *Physical review letters*, *120*(14), 14530.

[44] Patra, T. K., Meenakshisundaram, V., Hung, J. H., & Simmons, D. S. (2017). Neural-network-biased genetic algorithms for materials design: Evolutionary algorithms that learn. *ACS combinatorial science*, 19(2), 96-107.

[45] Hutchinson, M. L., Antono, E., Gibbons, B. M., Paradiso, S., Ling, J., & Meredig, B. (2017). Overcoming data scarcity with transfer learning. *arXiv preprint arXiv:1711.05099*.

[46] Kim, E., Huang, K., Saunders, A., McCallum, A., Ceder, G., & Olivetti, E. (2017). Materials synthesis insights from scientific literature via text extraction and machine learning. *Chemistry of Materials*, *29*(21), 9436-9444.

[47] Khorshidi, A., & Peterson, A. A. (2016). Amp: A modular approach to machine learning in atomistic simulations. *Computer Physics Communications*, *207*, 310-324

[48] Georgiou, T., Jalil, R., Belle, B. D., Britnell, L., Gorbachev, R. V., Morozov, S. V., ... & Eaves, L. (2013). Vertical field-effect transistor based on graphene–WS 2 heterostructures for flexible and transparent electronics. *Nature nanotechnology*, *8*(2), 100.

[49] Wang, Z., Ki, D. K., Chen, H., Berger, H., MacDonald, A. H., & Morpurgo, A. F. (2015). Strong interface-induced spin–orbit interaction in graphene on WS 2. *Nature communications*, *6*, 8339.

[50] Withers, F., Del Pozo-Zamudio, O., Mishchenko, A., Rooney, A. P., Gholinia, A., Watanabe, K., ... & Novoselov, K. S. (2015). Light-emitting diodes by band-structure engineering in van der Waals heterostructures. *Nature materials*, *14*(3), 301.

[51] Ross, J. S., Klement, P., Jones, A. M., Ghimire, N. J., Yan, J., Mandrus, D. G., ... & Cobden, D. H. (2014). Electrically tunable excitonic light-emitting diodes based on monolayer WSe 2 p–n junctions. *Nature nanotechnology*, *9*(4), 268.

[52] Wu, S., Buckley, S., Schaibley, J. R., Feng, L., Yan, J., Mandrus, D. G., ... & Xu, X. (2015). Monolayer semiconductor nanocavity lasers with ultralow thresholds. *Nature*, *520*(7545), 69.

[53] Ferrari, A. C., Bonaccorso, F., Fal'Ko, V., Novoselov, K. S., Roche, S., Bøggild, P., ... & Garrido, J. A. (2015). Science and technology roadmap for graphene, related two-dimensional crystals, and hybrid systems. *Nanoscale*, *7*(11), 4598-4810.

[54] Gogotsi, Y., & Anasori, B. (2019). The Rise of MXenes.

[55] Lin, Z., Carvalho, B. R., Kahn, E., Lv, R., Rao, R., Terrones, H., ... & Terrones, M. (2016). Defect engineering of two-dimensional transition metal dichalcogenides. *2D Materials*, *3*(2), 022002.

[56] Dabral, A., Lu, A. K. A., Chiappe, D., Houssa, M., & Pourtois, G. (2019). A systematic study of various 2D materials in the light of defect formation and oxidation. *Physical Chemistry Chemical Physics*, *21*(3), 1089-1099.

[57] Kang, K., Godin, K., Kim, Y. D., Fu, S., Cha, W., Hone, J., & Yang, E. H. (2017). Graphene- Assisted Antioxidation of Tungsten Disulfide Monolayers: Substrate and Electric- Field Effect. *Advanced Materials*, *29*(18), 1603898.

[58] Hynek, D. J., Pondick, J. V., & Cha, J. J. (2019). The development of 2D materials for electrochemical energy applications: A mechanistic approach. *APL Materials*, *7*(3), 030902.

[59] Frey, N. C., Wang, J., Vega Bellido, G. I., Anasori, B., Gogotsi, Y., & Shenoy, V. B. (2019). Prediction of Synthesis of 2D Metal Carbides and Nitrides (MXenes) and Their Precursors with Positive and Unlabeled Machine Learning. *ACS nano*, *13*(3), 3031-3041.

[60] Rajan, A. C., Mishra, A., Satsangi, S., Vaish, R., Mizuseki, H., Lee, K. R., & Singh, A. K. (2018). Machine-Learning-Assisted Accurate Band Gap Predictions of Functionalized MXene. *Chemistry of Materials*, *30*(12), 4031-4038.

[61] Zhao, H., Ezeh, C. I., Ren, W., Li, W., Pang, C. H., Zheng, C., ... & Wu, T. (2019). Integration of machine learning approaches for accelerated discovery of transition-metal dichalcogenides as Hg0 sensing materials. *Applied Energy*, *254*, 113651.

[62] Sendek, A. D., Yang, Q., Cubuk, E. D., Duerloo, K. A. N., Cui, Y., & Reed, E. J. (2017). Holistic computational structure screening of more than 12000 candidates for solid lithium-ion conductor materials. *Energy & Environmental Science*, *10*(1), 306-320.

[63] Wu, B., Han, S., Shin, K. G., & Lu, W. (2018). Application of artificial neural networks in design of lithium-ion batteries. *Journal of Power Sources*, *395*, 128-136.

[64] Kauwe, S. K., Rhone, T. D., & Sparks, T. D. (2019). Data-driven studies of li-ion-battery materials. *Crystals*, *9*(1), 54.

[65] Basu, S., Suresh, S., Ghatak, K., Bartolucci, S. F., Gupta, T., Hundekar, P., ... & Koratkar, N. (2018). Utilizing van der Waals slippery interfaces to enhance the electrochemical stability of Silicon film anodes in lithium-ion batteries. *ACS applied materials & interfaces*, *10*(16), 13442-13451.

[66] Zhang, C. J., Park, S. H., Seral- Ascaso, A., Barwich, S., McEvoy, N., Boland, C. S., ... & Nicolosi, V. (2019). High capacity silicon anodes enabled by MXene viscous aqueous ink. *Nature communications*, *10*(1), 849.

[67] Pomerantseva, E., & Gogotsi, Y. (2017). Two-dimensional heterostructures for energy storage. *Nature Energy*, *2*(7), 17089.

[68] Augustyn, V., & Gogotsi, Y. (2017). 2D materials with nanoconfined fluids for electrochemical energy storage. *Joule*, *1*(3), 443-452.